\newcommand{\kms}{{~\rm km\; s^{-1}}}
\newcommand{\cc}{{~\rm cm^{-3}}}
\newcommand{\msyr}{{~\rm M_{\odot}~yr^{-1}}}
\newcommand{\cm}{{~\rm cm}}
\newcommand{\s}{{~\rm s}}
\newcommand{\km}{{~\rm km}}
\newcommand{\g}{{~\rm g}}
\newcommand{\K}{{~\rm K}}
\newcommand{\erg}{{~\rm erg}}
\newcommand{\yr}{{~\rm yr}}
\newcommand{\Myr}{{~\rm Myr}}

\newcommand{\kpc}{{~\rm kpc}}

\documentclass[12pt,preprint]{emulateapj}
\usepackage{natbib}

\shorttitle{Fat bubbles}
\shortauthors{Sternberg}

\begin{document}

\title{INFLATING FAT BUBBLES IN CLUSTERS OF GALAXIES BY WIDE JETS}

\author{Assaf Sternberg\altaffilmark{1}, Fabio Pizzolato\altaffilmark{1},
Noam Soker\altaffilmark{1}}

\altaffiltext{1}{Department of Physics,
Technion$-$Israel Institute of Technology, Haifa 32000, Israel;
phassaf@techunix.technion.ac.il; fabio@physics.technion.ac.il;
soker@physics.technion.ac.il}

\begin{abstract}
We conduct two-dimensional hydrodynamical simulations of jets expanding in
the intra-cluster medium (ICM). We find that for a fat, i.e.
more or less spherical,
bubble attached to the center to be formed the jet should have high momentum
flux and a large opening angle.
Typically, the half opening angle should be $\alpha \ga 50^\circ$, and
the large momentum flux requires a  jet speed of
$v_j \sim 10^4 \km \s^{-1}$.
The inflation process involves vortices and local instabilities which mix
some ICM  with the hot bubble.
These  results predict that most of the gas inside the bubble has
a temperature of $3 \times 10^8 \la T_{b} \la 3 \times 10^{9} \K$, and that
large quantities of the cooling gas in cooling flow clusters are
expelled back to the intra-cluster medium, and heated up.
The magnetic fields and relativistic electrons that produce the
synchrotron radio emission might be formed in the shock wave of the jet.
\end{abstract}


\section{INTRODUCTION}
\label{sec:intro}

{\it Chandra} and {\it XMM-Newton} X-ray observations of clusters
of galaxies reveal the presence of X-ray-deficient bubbles in the
inner regions of many cooling flow clusters of galaxies, groups of
galaxies, and elliptical galaxies.
{{{ The hot  gas in the centers of many galaxies and clusters of galaxies
is a strong X-ray emitter, often this emission is strong enough to have a
dynamical role: the gas cools radiatively to very low temperatures
($T <10^4 \K$), and some  hotter  gas from regions further out
must flow inward to  keep the overall hydrostatic equilibrium in a
process dubbed {\em cooling flow}.
The high-resolution X-ray observations of {\em XMM--Newton} and
{\em Chandra} show that the amount of gas cooling to very low temperatures
is lower than the amount of hot gas which has a short cooling time.
This implies that some heating mechanism must be at work to
counterbalance the radiative losses.
It is widely agreed that a major role is played
by  the jets and bubbles formed by the active galactic nucleus
(AGN; see review by Peterson \& Fabian 2006) operating at the cluster
core.  Therefore, understanding the bubble formation is an
important key to understand  cooling flows. }}}
In some cases the
pairs of bubbles touch the center and are almost spherical
(``fat'' bubbles), with a typical hourglass shape (e.g., Perseus,
Fabian et al.\ 2000; A~2052, Blanton et al. 2003).

The basic condition for a jet to inflate a fat bubble is that
the jet's head will reside inside the bubble during the inflation phase.
For that to occur, it has been proposed that the either the jet's
opening angle is large (Soker 2004),
or the jet is narrow but its axis changes its direction.
The change in direction can result from precession (Soker 2004, 2007),
random change (Heinz et al. 2006), or a relative motion  between the ICM and the
AGN (Loken et al. 1995; Soker \& Bisker 2006; Rodriguez-Martinez et al. 2006).

As far as we know, none of the previous numerical simulations of jets expanding
from the center could inflate fat bubbles attached to the AGN.
{{{ The simulations by Basson \& Alexander (2003) come close to forming two fat,
almost spherical, bubbles, but the bubbles are too elongated. }}}
The need to understand the formation of fat bubbles by AGN jets was emphasized
in a recent meeting on cooling flows in galaxies and clusters
of galaxies (Pratt et al. 2007).

In this paper we report our success to inflate fat bubbles attached
to the center
by two-dimensional hydrodynamical numerical simulations.

\section{NUMERICAL METHOD AND SETUP}
\label{sec:numerics}

The simulations were performed using \emph{Virginia} \emph{Hydrodynamics-I}
(VH-1; Blondin et al. 1990 ; Stevens et al. 1992).
The code uses finite-difference techniques to solve the equations
of an ideal inviscid compressible fluid flow.
We used the two-dimensional version in spherical coordinates.
Namely, there is an azimuthal  symmetry, and the calculations
are performed in only one quarter of the meridional plane.
{{{ In the present paper we study the effect of spreading the ejected matter
over a large solid angle, namely, using a wide jet. For that purpose
a 2D code is adequate. More realistic 3D code will result in more
realistic fine-detail structures, but will not change the conclusions
regarding the conditions on inflating fat bubbles. }}}
Radiative cooling and gravity were not included, since  the total time of
the simulation, $<10^7 \yr$,  is somewhat shorter than the gravitational
time scale, and much shorter than the radiative cooling time. This
preliminary report aims  to emphasize the jet properties that
determines whether the required bubble is inflated,
and hence these omissions  are justified.

The initial density profile of the ICM is spherical, with a commonly used
(e.g., Vernaleo \& Reynolds 2006) profile of
\begin {equation}
\rho_{\rm ICM}= \rho_c [(1+(r/r_0)^2]^{-3/4}.
\label{rho}
\end {equation}
In the runs described here we take $\rho_c= 2.16 \times 10^{-25} \g\cc$,
and $r_0=100 \kpc$ (Rodriguez-Martinez et al. 2006). The ICM temperature is
$2.7 \times 10^7 \K$.
Since we study the inner region, the ICM density is practically constant
in the region we simulate $\rho \simeq \rho_c$. 

The jet is injected at a radius of $0.1 \kpc$, with constant mass flux and
constant radial velocity $v_j$ inside a half opening angle $\alpha$
(measured from the symmetry axis to the edge).
The total kinetic power of one jet is $\dot E_j=\dot M_j v_j^2/2$, where
$\dot M_j$ is the mass outflow rate in one jet.

\section{RESULTS}
\label{sec:results}

\begin{figure}  
\hskip 0.2 cm  
{\includegraphics[scale=0.30]{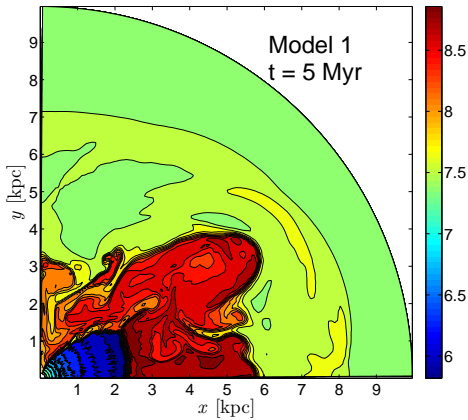}}
\caption{The temperature map of Model~1 at $t=5 \Myr$.
Scale is in $\log T(K)$.}
\label{temp}
\end{figure}

In Figure~\ref{opening} we show the density maps at two times and
for two jets with the same power and velocity (hence the same
mass outflow rate) but different  opening angles.
In Model~1 the half--opening angle is $\alpha=70^\circ$, the power of one jet
is $\dot E_j=10^{44} \erg \s^{-1}$, the initial jet's speed is
$v_j=7750 \kms$,
and hence the mass outflow rate in one jet in Model~1 is $5\msyr$.
Model~2 differs from Model~1 only in the half opening angle $\alpha=20^\circ$.
In Figure \ref{temp} we show the temperature map of Model~1 at $t=5 \Myr$.

Let us examine Figures \ref{opening} and \ref{temp} for Model~1 at $t=5\Myr$.
The very low density region close to the $x$-axis and the center is the region
where the jet is freely expanding,
Two shocks are seen in the figures. The forward shock
running through the ICM has an elliptical shape, extending in the figures
from $x=8.2 \kpc$ to $y=7.2 \kpc$. The reverse shock, where the
freely expanding jet is shocked, is seen as an almost vertical line at
$x\simeq 2.3 \kpc$.
\begin{figure}  
\vskip -0.5 cm
\vskip .5 cm  
{\includegraphics[scale=0.30]{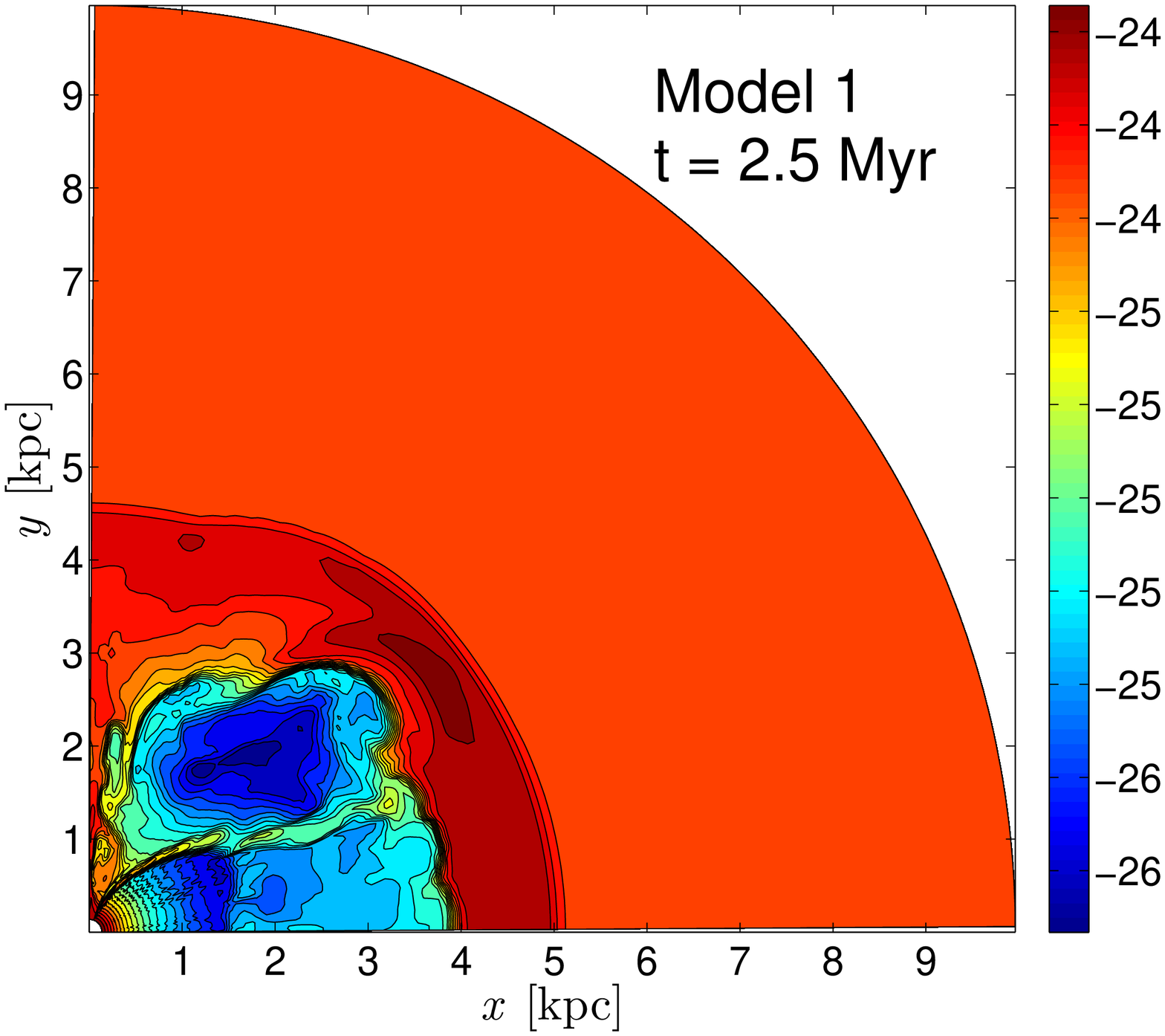}}  
{\includegraphics[scale=0.30]{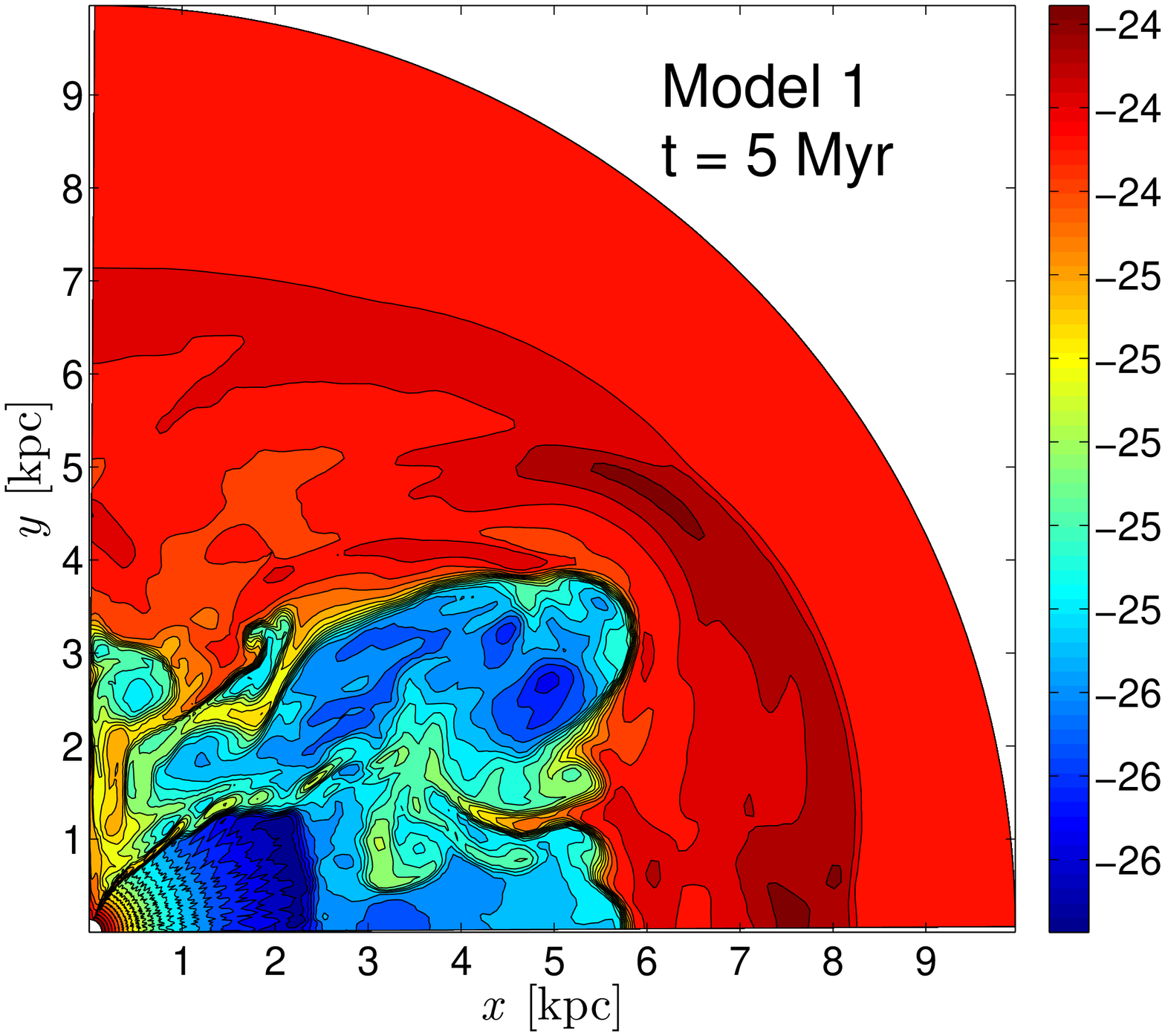}}  
{\includegraphics[scale=0.30]{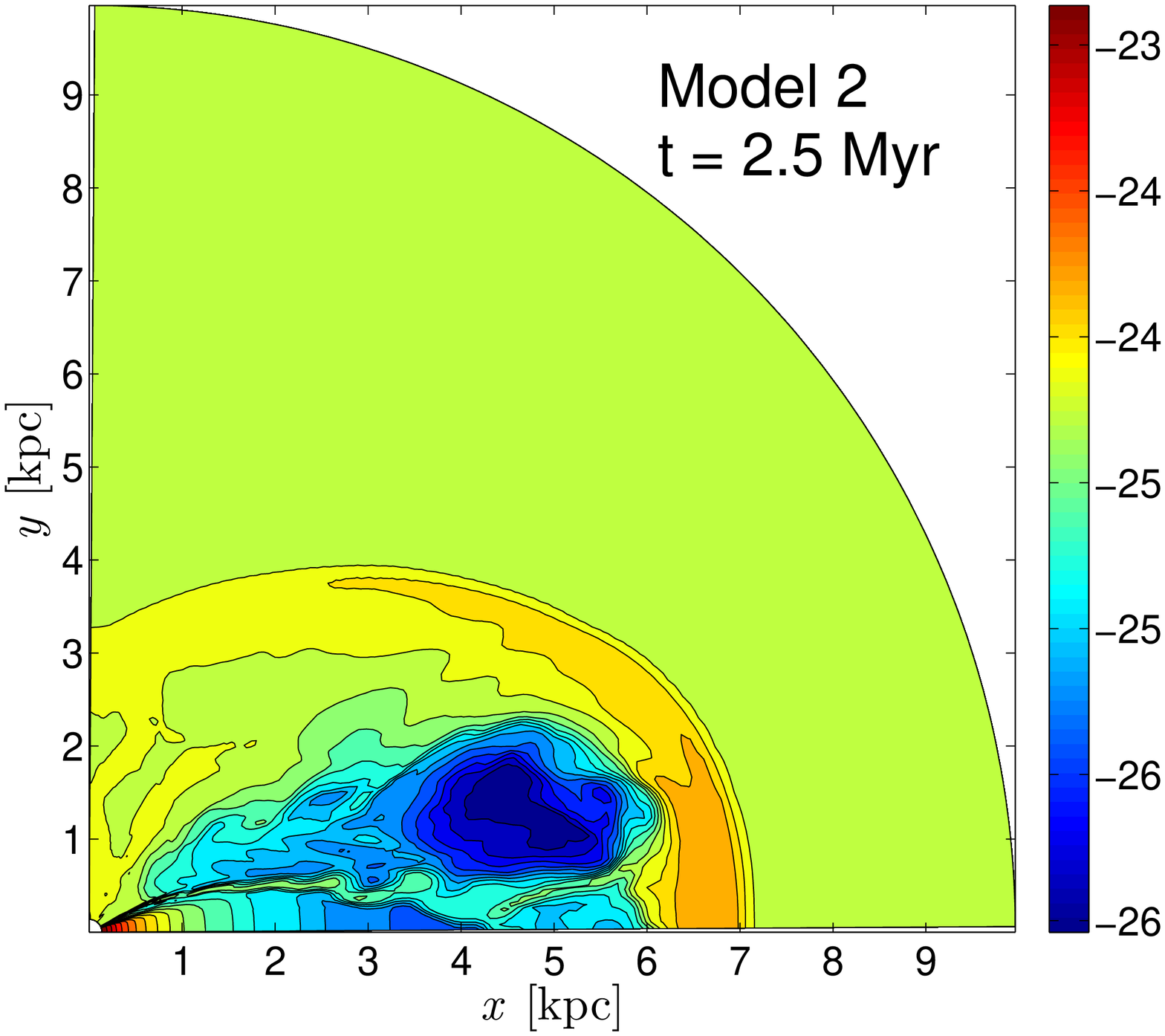}}
{\includegraphics[scale=0.30]{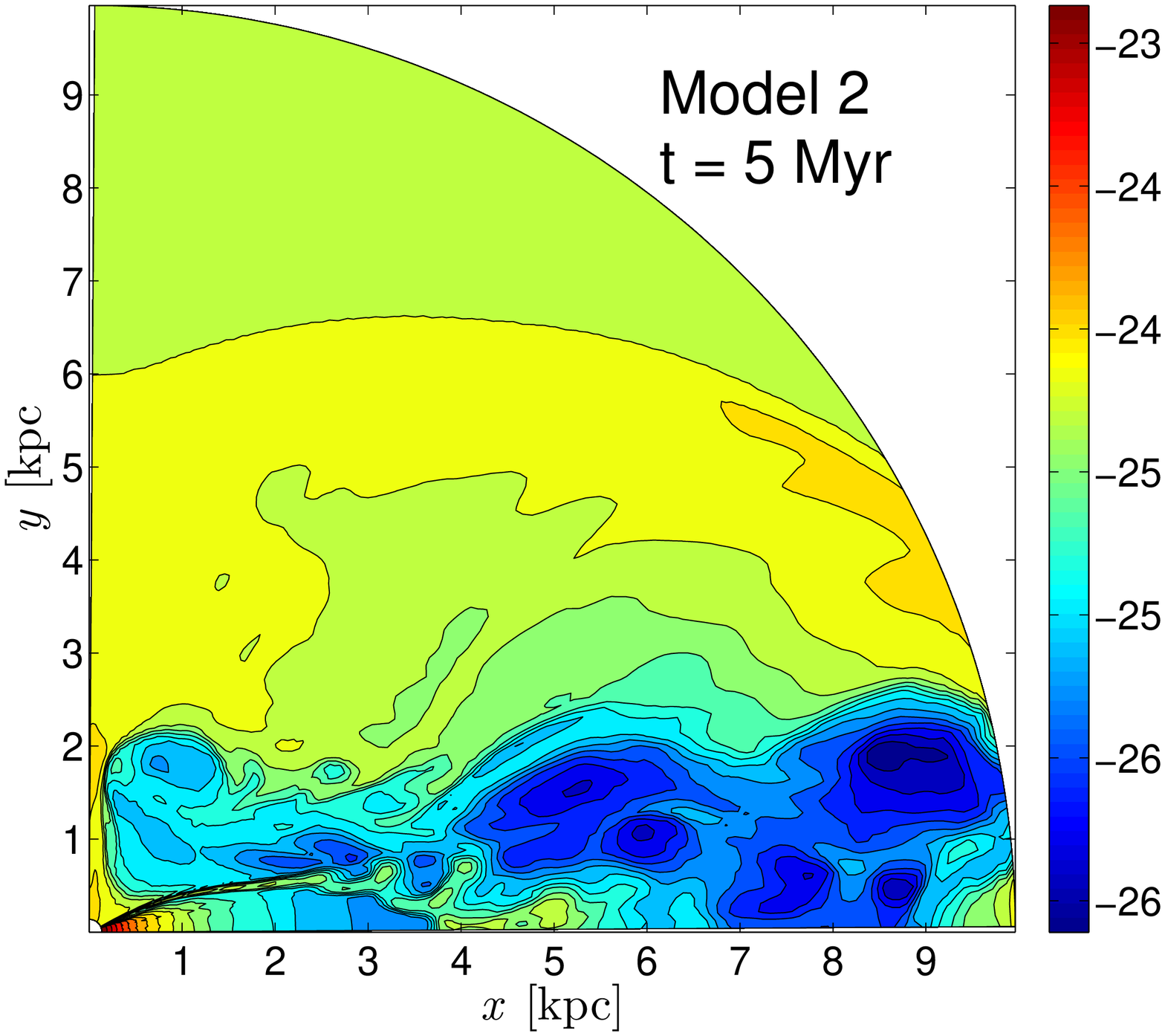}}
\end{figure}
\begin{figure}
\caption{Comparison of wide and narrow jets.
In Model~1 and Model~2 the power of one jet is $\dot E_j=10^{44} \erg \s^{-1}$,
and the initial jet's speed is $v_j=7750 \km \s^{-1}$.
The initial jet's half opening angle in Model~1 is $\alpha=70^\circ$,
and in Model~2 it is $\alpha=20^\circ$.
The horizontal lower edge is the symmetry axis, while the left side is
the equatorial plane; only quarter of the meridional plane is shown.
The density scale is on the right side of each panel (note the different scaling
between the panels), in $\log \rho ({\rm g} \cm^{-3})$. }
\label{opening}
\end{figure}

\begin{figure}  
\hskip -0.192 cm  
{\includegraphics[scale=0.30]{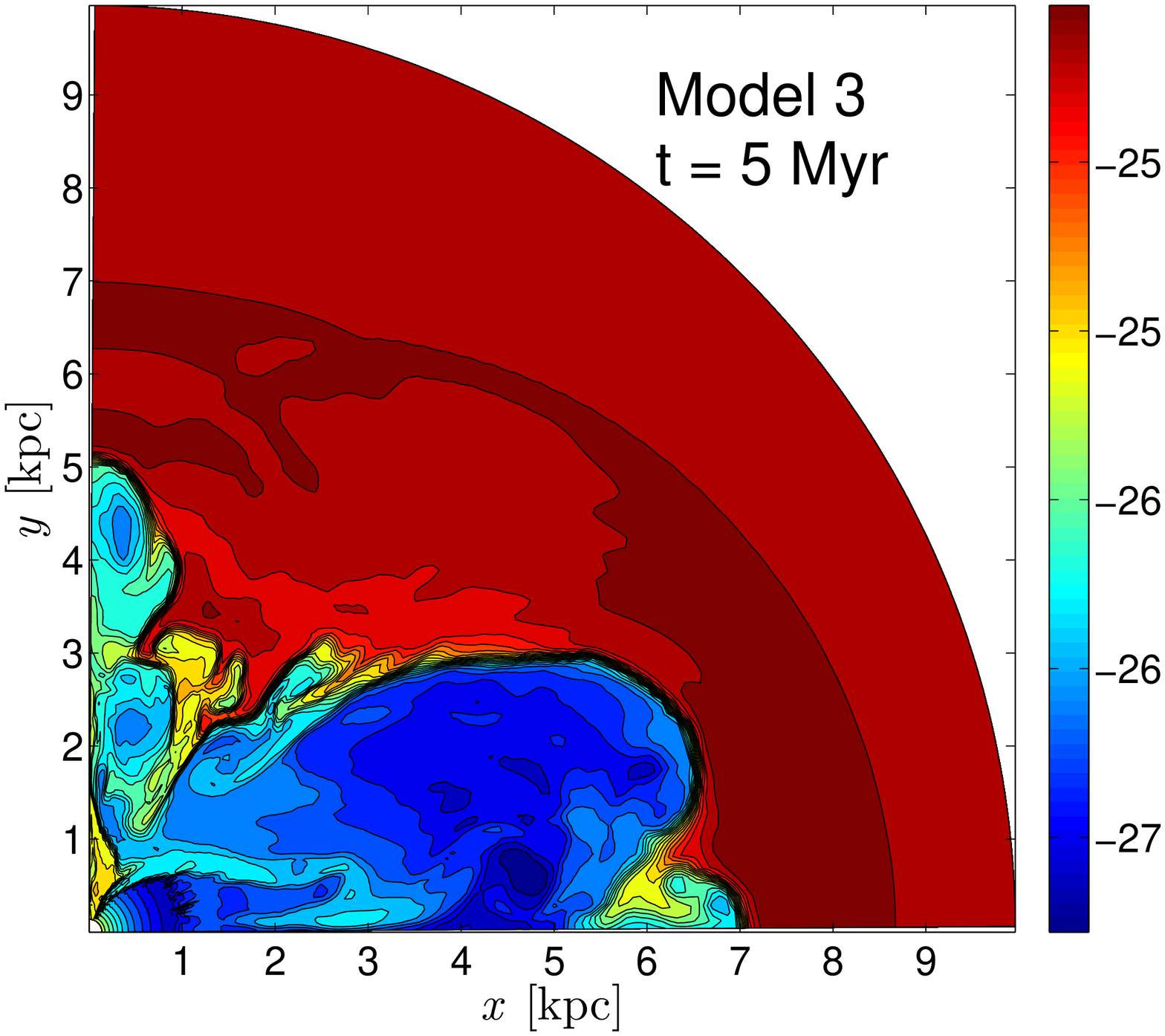}}
\hskip -1.7 cm  
{\includegraphics[scale=0.27]{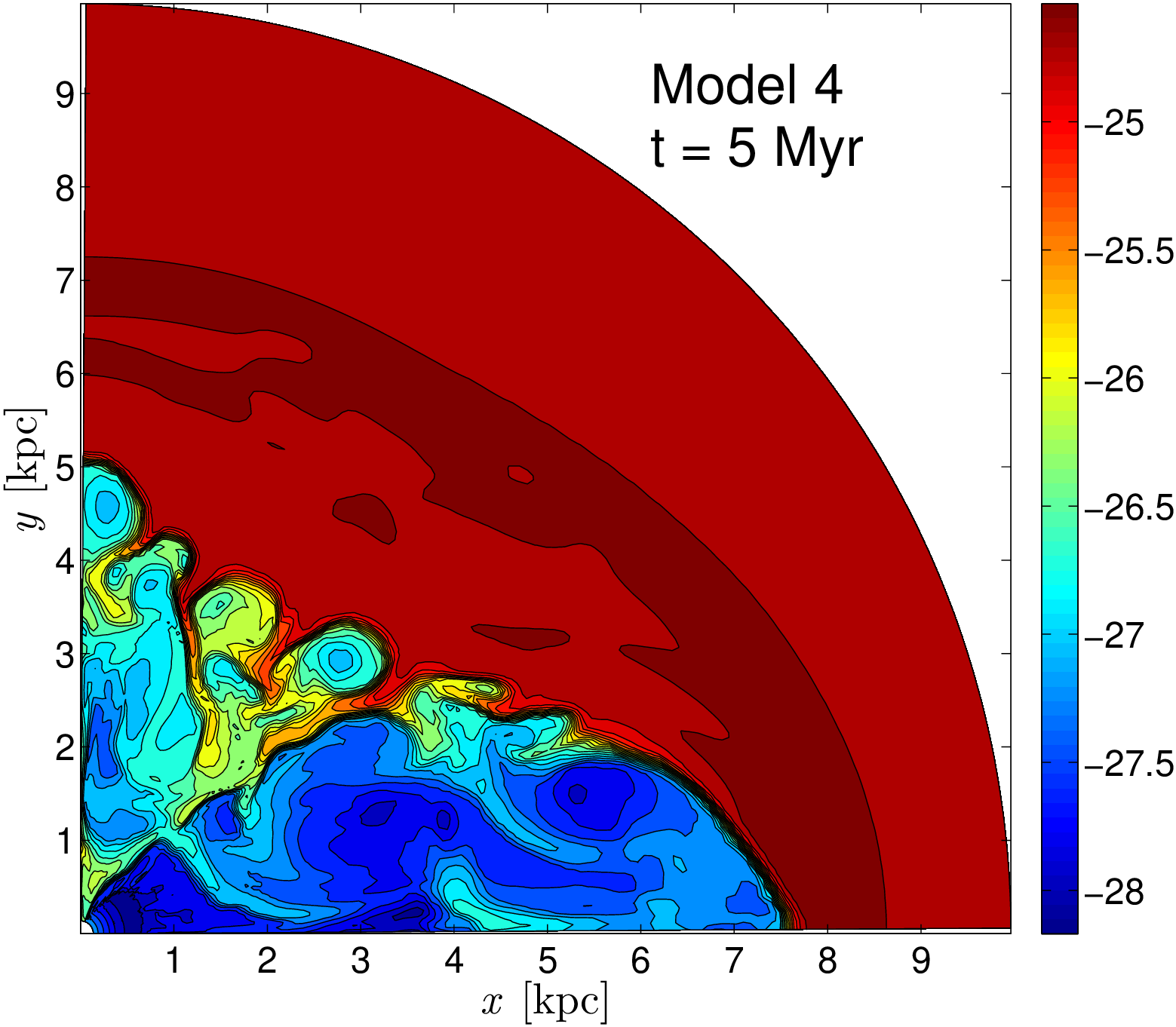}}
\hskip -2.2 cm  
{\includegraphics[scale=0.27]{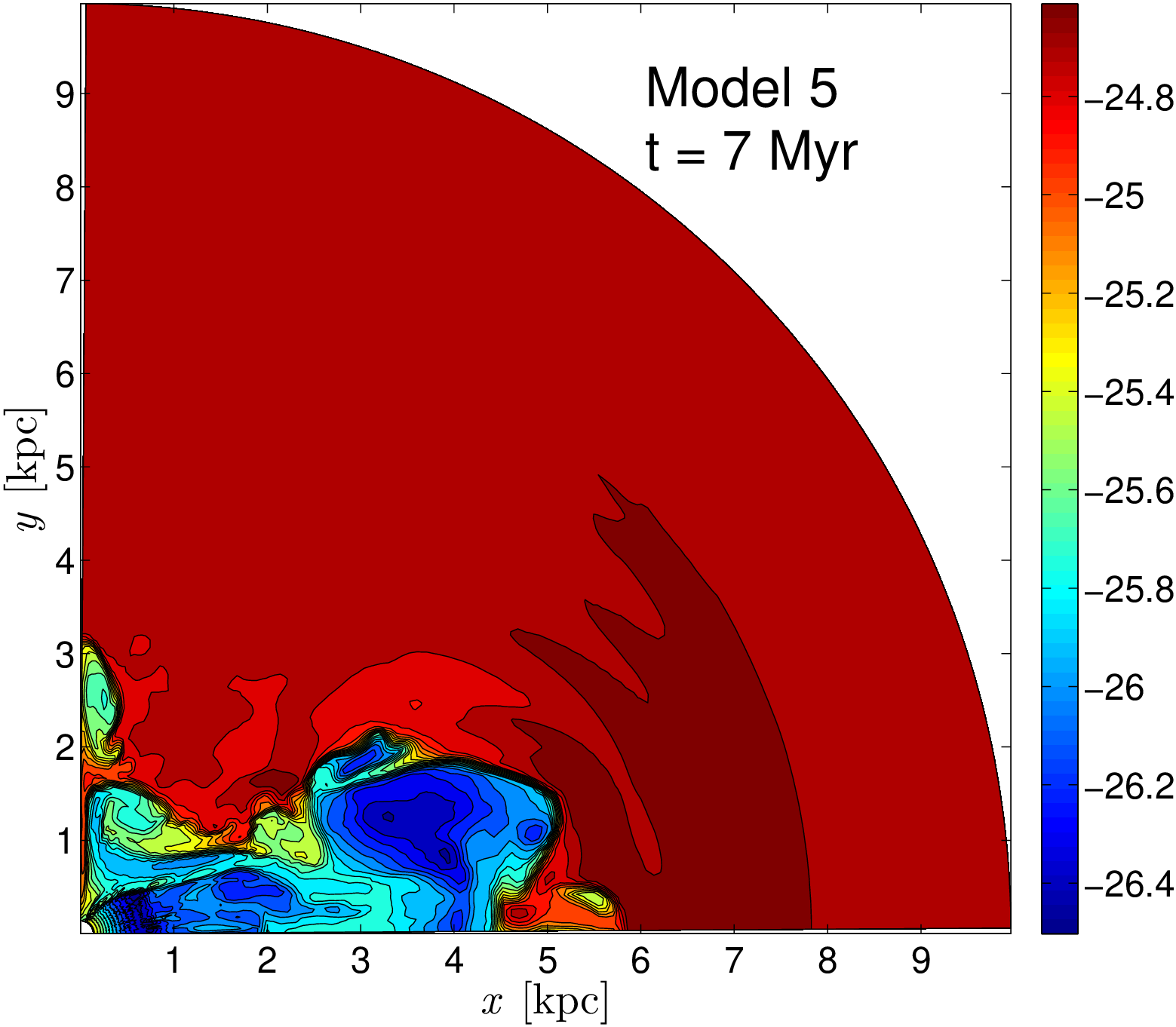}}
\caption{Same power and opening angle as in Model~1, but with jet speed
3 times (Model~3) and 9 times (Model~4) larger.
Note that the reverse shock in Model~3 (4) is located at $r\sim 0.9 \kpc$ (0.6 \kpc).
Model~5 has the same jet speed as Model~1 but a power lower by a factor of 10. }
\label{parameters}
\end{figure}

Model~3 has the same jet's power and opening angle, but it differs
from Model~1 in  having a jet faster  by a factor of $3$
($v_j=2.32 \times 10^4 \km \s^{-1}$),
and hence a mass outflow rate lower by a factor of $9$.
Model~4 has the same jet's power, but it differs from Model~1 in having
a faster jet,
by a factor of $9$, of $v_j=6.97 \times 10^4 \km \s^{-1}$, and hence a
mass outflow rate $\dot M_j=0.065 M_\odot \yr^{-1}$,
 lower by a factor of $81$ than that in Model~1.
In Model~5 the jet's speed is as in Model~1, but the power and
mass outflow rate are
lower by a factor of 10, i.e., $\dot E_j=10^{43} \erg \s^{-1}$.
The density maps of these jets are presented in Figure \ref{parameters},
at times as indicated in each panel.
%

As evident by Figures \ref{opening} and \ref{parameters} fat bubbles
attached to the center are formed only with wide ($\alpha \ga 50^\circ$)
and slow
($v_j \simeq 5,000-30,000 \kms$) jets, as in Models~1 and Model~3.
The case of model~5 is marginal. The jet is too weak to blow
a large fat bubble. More energetic jets, for the ambient medium
used by us, can
blow fat bubbles, if they wide and slow enough.
We also run a model where both the ambient density and the mass outflow rate,
hence power,
were $0.33$ times their values in Model~1 (this is more
appropriate to the case in Perseus). Beside small numerical differences,
the results are the same as in Model~1.
Therefore, models with a power of $10^{42}-10^{43} \erg \s^{-1}$
can still blow fat
bubbles if the ambient medium density is proportionally lower than
what  used here in Model~1.

In all panels we see the vortices on the sides of the jet,
developed by the back-flowing jet's material (termed ``cocoon''),
and Kelvin-Helmholtz instabilities (Norman et al. 1982).
It is   well--known (e.g., Loken et al. 1992) that
narrow jets  interacting with their environment
form cylindrically shaped cocoons
(e.g., Krause 2004, 2005; O'Neill et al. 2004).

The structure of the cocoon and its width depends on the Mach number and
density contrast between the jet and the environment as well (Krause 2003).
Rosen et al. (1999) found that relativistic jets have small cocoons, while
slow jets develop  large cocoons. Therefore, lower velocities increase
the cocoon size in jets, as we also find here.
However, the simulations cited above did not lead to the formation of
fat bubbles attached to the center,

The wide jet's angle used here for the slow jets, lead to the desired
morphology, containing the following components:
\begin{enumerate}
\item A fat, very low density cavity.
\item A dense shell surrounding the cavity
from all directions, including in the equatorial plane (adding gravity
might enhance the dense shell concentration there).
\item A slow shock of the dense shell. The average
speed of the forward shock between $t=2.5 \Myr$ and $t=5 \Myr$ in Model~1
is $\sim 1400 \kms$, which corresponds to a Mach number of $\sim 2$.
\end{enumerate}
Inclusion of magnetic fields in the simulations might reduce
the instability, making a
smoother boundary for the bubble. Inclusion of heat conduction could lead to
the evaporation of denser filaments penetrating into the bubble.

Soker's (2004) formula predicts that wider jets will more easily form fat
bubbles, as we find here. However, that formula predicts also that faster
jets are more likely to inflate fat bubbles, contrary to our finding.
The reason for the discrepancy is that Soker (2004) assumed that the jet has
already reached a large distance of few kpc or more from the center.
This is the case in the simulations conducted by Vernaleo \& Reynolds (2006),
where bubbles were inflated at a distance of $400 \kpc$ from the center,
agreeing with the estimate of Soker (2004; see Soker 2007).
However, for a given jet's kinetic power a higher velocity implies a
lower momentum flux,
and hence the jet is recollimated by the pressure formed as it
interacts with the ICM.
This is clearly seen in Model~4, which starts with a wide angle,
but  is soon recollimated to a narrow jet that expands to a
large distance without forming a fat bubble.

\section{DISCUSSION AND SUMMARY} \label{sec:discussion}

We showed that slow wide jets (Model~1 and Model~3) can inflate fat bubbles
attached to the center.
Morphologically, these resemble the structures observed in some cooling
flow clusters like  A~2052 and Perseus.
This by itself does not imply that these bubbles are inflated by wide jets,
as there is still the possibility that these are inflated by
jets with  varying axis (Soker 2004, 2007; Heinz et al. 2006).

Our most controversial finding might be that the jet should be
highly sub-relativistic
(but still of high Mach number), with a typical velocity of
$v_j \sim 10^4 \kms$.
This leads to a bubble temperature of $T_b \simeq 3 \times 10^8 - 3 \times 10^9 \K$.
Our motivation and justifications for using relatively
slow massive jets are as follows.
\begin{enumerate}
\item
Equipartition arguments applied to the bubble systems in several cooling
flow clusters
suggest that the pressure in the radio lobes is an order of magnitude less than
the surrounding thermal gas pressure (e.g., Hydra~A, McNamara et al. 2000;
Perseus, Fabian et al. 2000; A~2052, Blanton et al. 2001).
Therefore, as noted by these authors, a likely solution is that there is
a good amount of thermal pressure in the bubbles, resulting from gas at
several tens of keV {{{ (several$\times10^8 \K$.) }}}
Sanders \& Fabian (2007) constrain the amount  of
hot thermal plasma in the temperature range $7-70 \times 10^7 \K$
in the bubbles of Perseus. For  this cluster, their
result is consistent with our
model provided that the jet speed is
$v_j \ga 2 \times 10^4 \kms$.
\item
Nearby Seyfert galaxies have slow outflows of $\sim 1000\kms$
{{{ (e.g., Crenshaw \& Kraemer 2007). }}}
It has been claimed (Behar et al. 2003) that these are the
large-angle high mass loss rate winds observed directly in Seyfert 2's.
There are also claims for faster winds in brighter quasars
($24000 \kms$, Pounds et al. 2003;   but see
Kaspi \& Behar 2006) for which mass
estimates are  still unavailable.
Possibly, UV broad absorption line quasars (BALQS)
may be related to the winds simulated in this work, although they have
not been observed in any of the known cooling flow clusters and their
entrained mass is unknown.
\item
On the theoretical side, Binney (2004) argued that in many cases the
relativistic jet carries a
small fraction of the mass and energy in the outflow.
In their jet simulations Omma et al. (2004) take $v_j = 10^4 \kms$ and
a jet mass outflow rate of $2 \msyr$.
There are other models which predict slow ($\sim 10^4 \kms$) winds
(e.g., Begelman \& Celotti 2004).
\item
Soker \& Pizzolato (2005) proposed a scenario in which a large fraction,
or even most, of the gas cooling to low temperatures of $T < 10^4 \K$ in
cooling flow clusters directly gains energy from the central black hole.
In Model~1, the mass outflow rate in each jet is $5 \msyr$,
or $10 \msyr$ in the two jets. Therefore, it is possible that
the inflation of fat attached bubbles occur when mass outflow rate is so high,
that it must occur in cooling flow clusters.
\end{enumerate}
Dunn et al. (2006) argue that the jets in Perseus
(a cluster with a clear indication of radio bubbles)
are dominated by electron-positron plasma.  This result might be
controversial, as other authors (Celotti \& Fabian, 2003) prefer
baryon-dominated, not lepton-dominated jets.
Anyhow, even a lepton-dominated jet would not be at odds
with our claim that the bulk of the energy is thermal.
The conclusions of Dunn et al. (2006) apply at the base of the
jet (on parsec scales), which does not exclude a baryonic contamination of
the outflow further out.
In the present framework, therefore, the bulk of the energy input by the
jet may be thermal, with $\sim 10\%$ of the energy budget in  magnetic
fields and relativistic particles, in agreement with the
observations of the cluster' cavities in the radio band (e.g.
Clarke et al. 2003 and references therein).

If the acceleration of electrons in ICM shocks is as efficient as that in
supernova remnant shocks of few~$10^3 {\rm \km \s}^{-1}$ as suggested
by Loeb \& Waxman (2000), then we can expect~$\sim 5\%$ of the post
shock energy to be converted to relativistic electrons and magnetic fields.
This is sufficient to account for the radio synchrotron emission inside bubbles
(B{\^i}rzan et al. 2004, and references therein).


\acknowledgements
We thank John Blondin for his great help with the numerical code.
This research was supported by the Asher Fund for Space Research at the
Technion. It is a pleasure to thank Ehud Behar for  useful discussions.

\end{document}